\documentclass[preprint,authoryear, 3p]{elsarticle}

\usepackage{amssymb}
\usepackage{amsmath}
\usepackage{subfig,graphicx}
\usepackage{ulem}

\usepackage[breaklinks=true, colorlinks=true, urlcolor=magenta, citecolor=blue, linkcolor=blue]{hyperref}

\newcommand{\orb}{${\Omega}$\,} 
\newcommand{\twoorb}{${2\Omega}$\,} 
\newcommand{\spin}{${\omega}$\,} 
\newcommand{\twosp}{${2\omega}$\,} 
\newcommand{\be}{${\omega-\Omega}$\,} 
\newcommand{\twobe}{${2(\omega-\Omega)}$\,} 
\newcommand{\porb}{P$_{\Omega}$\,} 
\newcommand{\psp}{P$_{\omega}$\,} 
\newcommand{\pbe}{P$_{\omega-\Omega}$\,} 

\journal{New Astronomy}

\begin{document}

\begin{frontmatter}

\title{Long-term Optical Photometry of V709 Cas Using \textit{TESS}:  Refined Periods and Accretion Geometry} 

\author[a,b]{Srinivas M Rao} 
\ead{srinivas@aries.res.in}
\author[a]{Jeewan C Pandey}
\ead{jeewan@aries.res.in}
\author[a,c]{Nikita Rawat}
\author[d]{Arti Joshi}
\author[e]{Ajay Kumar Singh}

\affiliation[a]{,
            addressline={Aryabhatta Research Institute of Observational Sciences (ARIES)}, 
            city={Nainital},
            postcode={263001}, 
            state={Uttarakhand},
            country={India}}

\affiliation[b]{,
            addressline={Mahatma Jyotiba Phule Rohilkhand University}, 
            city={Bareilly},
            postcode={243006}, 
            state={U.P.},
            country={India}}

\affiliation[c]{,
            addressline={South African Astronomical Observatory, PO Box 9, Observatory}, 
            city={Cape Town},
            postcode={7935}, 
            state={},
            country={South Africa}}

\affiliation[d]{,
            addressline={Institute of Astrophysics, Pontificia Universidad Católica de Chile, Av. Vicuña MacKenna 4860}, 
            city={Santiago},
            postcode={7820436}, 
            state={},
            country={Chile}}

\affiliation[e]{,
            addressline={Department of Applied Physics/Physics, Bareilly College}, 
            city={Bareilly},
            postcode={243006}, 
            country={India}}
            
\begin{abstract}
Utilizing high-cadence and long-term optical photometry from the Transiting Exoplanet Survey Satellite (\textit{TESS}), we conducted a time-resolved timing analysis of the Intermediate Polar, V709 Cas. Our analysis reveals key characteristics of this system: an orbital period of 5.3329 $\pm$ 0.0002 h, a spin period of 312.7478 $\pm$ 0.0002 s, and a beat period of 317.9267 $\pm$ 0.0002 s.  These periods represent a significant refinement over the results of previous studies.  These analyses demonstrate that V709 Cas is primarily a disc-overflow accretor, with accretion predominantly occurring via a disc. However, the analysis also reveals epochs where stream-fed accretion is a dominant accretion process.  Time-resolved analysis of 20 s short cadence data obtained from sectors 57 and 58 reveals the presence of distinct first harmonics of the spin and beat frequencies.  This finding indicates the presence of double peak spin modulation, a characteristic signature of two-pole accretion onto the white dwarf.
\end{abstract}

\begin{keyword}
Cataclysmic variable stars \sep DQ Herculis stars \sep Stellar accretion \sep accretion disc \sep binaries \sep star: V709 Cas 



\end{keyword}

\end{frontmatter}



\section{Introduction}
\label{sec:intro}
Cataclysmic Variables (CVs) are semi-detached interacting binary systems characterized by the dynamic interaction between a white dwarf (WD), referred to as the primary, and a late-type main sequence star, known as the secondary, which overflows its Roche lobe, leading to material accretion onto the WD. These systems are categorized into two distinct groups based on the strength of the WD's magnetic field: non-magnetic CVs (NMCVs) and magnetic CVs (MCVs). NMCVs are identified by their relatively weak magnetic fields, typically measuring less than 0.1 million Gauss (MG), which results in the magnetic field having minimal influence on the accretion process from the secondary star to the WD \citep{Warner1986}. In contrast to NMCVs, MCVs are characterized by significantly stronger magnetic fields. The strong magnetic field in MCVs plays a pivotal role in governing the accretion processes within these systems.  Based on the strength of the magnetic field and, thus, the accretion mechanism, MCVs are classified into two subclasses: polars and intermediate polars (IPs). Polars are unique for their exceptionally powerful magnetic fields, exceeding 10 MG \citep{Cropper1990}. On the other hand, IPs are thought to exhibit magnetic field strength, typically ranging from 1 to 10 MG \citep{Patterson1994}.  

Three distinct accretion mechanisms, namely disc-fed, stream-fed, and disc-overflow, are generally observed in IPs. In the disc-fed mechanism, the accretion process occurs from a Keplerian disc, and this disc is disrupted at a specific point known as the magnetospheric radius of the WD \citep[e.g.][]{Hellier1989a}. Within the stream-fed mechanism, the material is directly accreted onto the surface of the WD by following the magnetic field lines \citep{Hameury1986}. However, in the disc-overflow mechanism, some fraction of the material overflows from the accretion disc and collides with the magnetosphere. Consequently, both modes of accretion, stream-fed and disc-fed, operate concurrently as described in \citet{Hellier1989b}. An additional conceivable accretion mechanism to consider is diamagnetic blob accretion. Within this mechanism, both stream-fed and disc-fed processes can coexist, but the accretion flow is characterized by diamagnetic clumpy structures \citep{Lubow1989,king1993}.

The accretion flow in the several IPs is found to be variable using the long-term X-ray
and optical observations and switches between disc-dominance and stream-dominance accretion. V2400 Oph, FO Aqr, TX Col, and V902 Mon are examples of a few such IPs that show the change in their accretion flows, switching from disc dominance to stream dominance and vice versa \citep{Joshi2019, Littlefield2020, Rawat2021, Rawat2022, Pandey2023}. Consequently, this paper focuses on analyzing the accretion flow in V709 Cas, an intermediate polar with a spin-to-orbital period ratio below 0.1, utilizing short-cadence, long-baseline \textit {TESS} observations. This study represents the first analysis of continuous long-term and short-cadence observations of V709 Cas, uniquely enabled by \textit{TESS} data.

V709 Cas (=RX J0028.8+5917) was identified as an IP by \cite{HaberlMotch1995} in the \textit{ROSAT} all-sky survey with spin (\psp) and orbital (\porb) periods of 312.78 s and  5.332879 h, respectively \citep{Norton1999, Motch1996, Thorstensen2010}. In addition to \psp, \cite{Kozhevnikov2001}  have detected the optical beat period (\pbe) of 317.94 s, whereas \cite{deMartino2001} have shown the first harmonic of beat frequency (\be). Using various methods, the mass of WD is estimated to be in the range of 0.6 - 1.2 M$_\odot$ 
\citep{Ramsay2000, Suleimanov2005, Falanga2005, Yuasa2010, Shaw2018}.

The paper is organized as follows. In section \ref{sec:obs}, we describe observations and data. Section \ref{sec:ana} contains the analysis and the results. Finally, we
present the discussion and summary in sections \ref{sec:dis} and \ref{sec:conc}, respectively.

\begin{table*}
    \centering
    \caption{Observation log of V709 Cas.}
    \begin{tabular}{l c c c}
    \hline
     Sector & Start Time & End Time & No. of observing days\\
    \hline
     ~~17 & 2019-10-08 04:28:21 & 2019-11-02 04:20:45 & 24.99\\
     ~~18 & 2019-11-03 03:48:46 & 2019-11-27 12:24:09 & 24.36\\
     ~~24 & 2020-04-16 06:58:02 & 2020-05-12 18:38:01 & 26.49\\
     ~~57* & 2022-09-30 20:33:19 & 2022-10-29 14:50:01 & 28.76\\
     ~~58* & 2022-10-29 19:58:01 & 2022-11-26 13:13:26 & 27.72\\ 
     ~~78 & 2024-05-03 18:24:15 & 2024-05-21 19:38:36 & 18.05\\
    \hline
    \end{tabular}
    ~~\\
    \text{*} The data is also available for the 20 s cadence.
    \label{tab:obslog}
\end{table*}

\addtolength{\tabcolsep}{-2pt}
\begin{table*}[ht]
    \centering
    \caption{Periods obtained corresponding to the dominant peaks of the LS power spectra.}
    \label{tab:ps}
    \begin{tabular}{lccccc}
    \hline
      Sector &  $P_\Omega$(h) & $P_\omega$(s) & $P_{\omega-\Omega}$(s) & $P_{2\omega}$ (s) & $P_{2(\omega-\Omega)}$ (s)\\
     \hline
      17 & $5.34 \pm 0.01$ & $312.75 \pm 0.01$ & $317.94 \pm 0.01$ & - & -\\
      18 & $5.33 \pm 0.01$ & $312.75 \pm 0.02$ & $317.92 \pm 0.01$ & - & -\\
      24 & $5.33 \pm 0.01$   & $312.74 \pm 0.01$ & $317.93 \pm 0.01$ & - & -\\
      57 & $5.33 \pm 0.01$ & $312.75 \pm 0.01$ & $317.93 \pm 0.01$ & - & -\\
      57$^*$ & $5.33 \pm 0.01$ & $312.74 \pm 0.01$ & $317.92 \pm 0.01$ & $156.374 \pm 0.002$ & $158.962 \pm 0.002$\\
      58 & $5.32 \pm 0.01$ & $312.74 \pm 0.01$ & $317.92 \pm 0.01$ & - & -\\
      58$^*$ & $5.33 \pm 0.01$ & $312.75 \pm 0.01$ & $317.93 \pm 0.01$ & $156.374 \pm 0.003$ & $158.964 \pm 0.003$\\
      78 & $5.35 \pm 0.02$ & $312.74 \pm 0.02$ & $317.90 \pm 0.02$ & - & -\\
      Combined$^\dagger$ & $5.3329 \pm 0.0002$ & $312.7478 \pm 0.0002$ & $317.9267 \pm 0.0002$ & - & -\\
    \hline
    \end{tabular}\\
{$^*$ for 20 s cadence\\
$\dagger$ Represents the periods derived from the power spectra of combined \textit{TESS} observations of all sectors.} 
\end{table*}

\section{Observations and Data}
\label{sec:obs}
We have used the data obtained from high cadence long-baseline \textit{TESS} observations. The \textit{TESS} \citep{Ricker2015} instrument consists of four wide-field CCD cameras that can image a region of the sky, measuring $24^\circ \times 96^\circ$. \textit{TESS} observations are divided into sectors, each lasting two orbits, or about 27.4 days. \textit{TESS} bandpass extends from 600 nm to 1000 nm with an effective wavelength of 800 nm. The data is stored in the Mikulski Archive for Space Telescopes data\footnote{\url{https://mast.stsci.edu/portal/Mashup/Clients/Mast/Portal.html}} with unique identification number `TIC 320180973'. We have used the Pre-search Data Conditioned Simple Aperture Photometry (PDCSAP) flux, which is the Simple Aperture Photometry (SAP) flux values corrected for instrumental variations\footnote{See section 2.1 of the TESS archive manual at \url{https://outerspace.stsci.edu/display/TESS/2.1+Levels+of+data+processing}}. PDCSAP flux also corrects for flux captured from the nearby stars. Data from an anomalous event had quality flags greater than 0 in the fits file. We have considered the PDCSAP flux data for which the quality flag is marked as `0'. The log of observations is shown in Table \ref{tab:obslog}. Additionally, 20 s cadence data were acquired for two sectors 57 and 58.

\begin{figure*}
    \centering
    \subfloat[Daily averaged light curve of V709 Cas from all sectors of \textit{TESS}.]{ \includegraphics[width=0.95\textwidth]{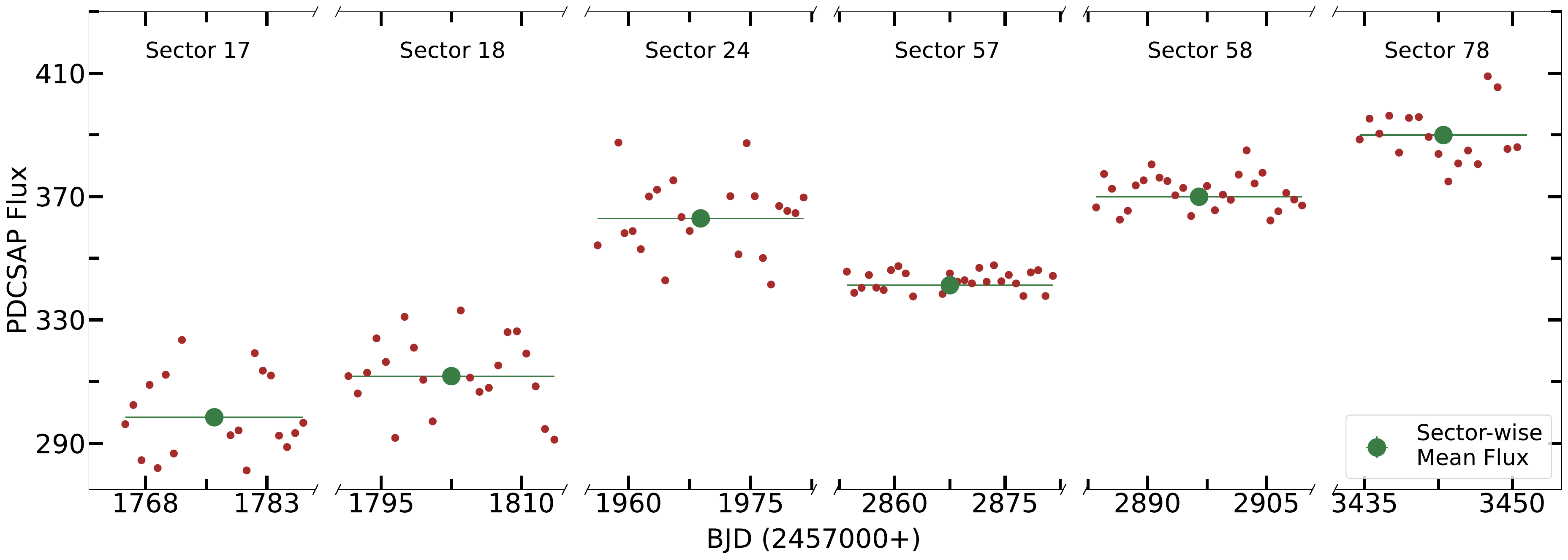}\label{fig:mean_lc}}
    
    \subfloat[A three days representative light curve from sector 17.]{ \includegraphics[width=0.95\textwidth]{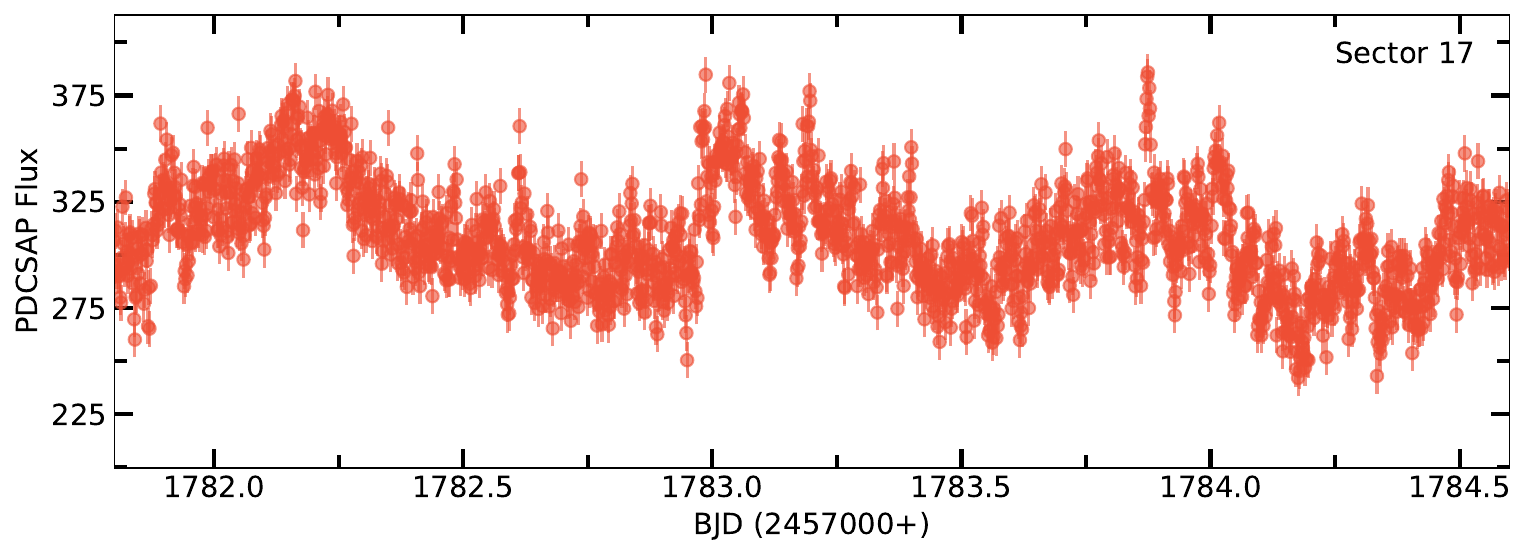}\label{fig:lc}}
\caption{(a) The daily averaged light curve of V709 Cas.  The green solid circles show the mean flux of the individual sector, and the error bar on the x-axis shows that sector's start and end times. (b) Three days representative \textit {TESS} light curve of V709 Cas from sector 17 observations. }

    \label{fig:v709cas_lc_ps}
\end{figure*}

\section{Analysis and Results}
\label{sec:ana}
\subsection{Light curves and power spectra}
\begin{figure*}
    \centering
    \includegraphics[width=\textwidth]{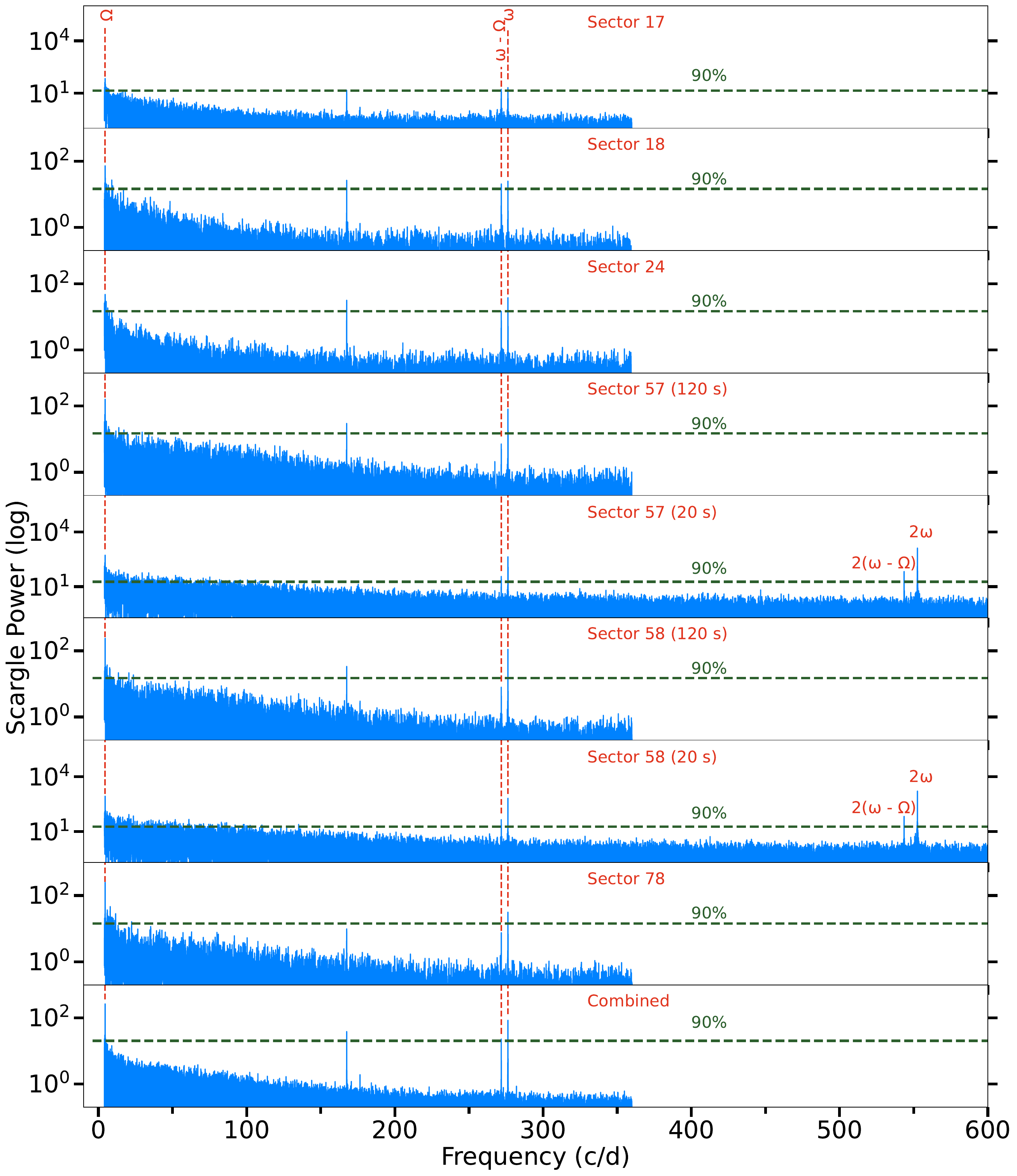}
    \caption{Power spectra were obtained for all the sectors and the combined dataset. The identified different frequencies are marked by vertical red dashed lines. The green horizontal dashed line represents the 90\% confidence level.}
    \label{fig:v709cas_ps}
\end{figure*}

The day-averaged light curve across all \textit{TESS} sectors is presented in Figure \ref{fig:mean_lc} whereas  Figure \ref{fig:lc} shows the light curve for three consecutive days from the observations of sector 17 for V709 Cas. The averaged light curve shows the variation in the daywise time scale for sectors 17, 18, 24 and 78, whereas a consistent averaged flux for each day was observed for sectors 57 and 58. The light curve of V709 Cas also showed the variations in a short time scale too, as shown in Figure  \ref{fig:lc}. The time series data from all six sectors were separately analysed using the Lomb-Scargle (LS) periodogram \citep{Lomb1976, Scargle1982} method to know the periodic behaviour of the light curve. The power spectra of all sectors' combined data set, along with the power spectra of individual sectors, are shown in Figure \ref{fig:v709cas_ps}.
We have detected several dominant peaks in each sector's power spectra. The significance of these detected peaks was determined by calculating the false-alarm probability (FAP; \citealt{Horne1986}), which is represented by a grey surface for the significance level of 90$\%$. The frequencies identified above the significance level are marked with a black dashed line.

We have detected three dominant peaks at the orbital (\orb), spin (\spin), and beat (\be) frequencies in the power spectra of different sectors. However, due to better time resolution for sectors 57 and 58, harmonics of both \spin and \be frequencies were also evident in the power spectra of 20 s cadence data. The periods and associated errors corresponding to these frequencies are given in Table \ref{tab:ps}. Errors were derived by calculating the half-size of a single frequency bin, centred on the peak in the power spectra. A cluster of frequencies in the lower frequency region of 0-4 cycle d$^{-1}$ was also found to be present in the power spectrum with no obvious relation to the frequencies identified in the power spectrum. Also, these frequencies are inconsistent among different sectors. Therefore, we have not shown this region of frequencies in the presented power spectrum, which could be probably due to long-term trends in the data. 
The \orb and \spin frequencies were found to be present in all sectors above the FAP level, whereas \be frequency was found to be present above the FAP level for all sectors excluding sector 78. We noted that the \be frequency was below the FAP level in the 120 s cadence data of sectors 57 and 58 but above the FAP level in the 20 s cadence data. The \twoorb frequency was also present for the sectors 18 and 57. The combined power spectra also show the presence of \orb, \twoorb, \be, and \spin frequencies. The first harmonics of \spin and \be were stronger than the fundamental frequencies in the short cadence data of sectors 57 and 58.  

\begin{figure*}
    \centering
    \subfloat[Power spectra of sectors 57 and 58]{\includegraphics[width=0.47\textwidth]{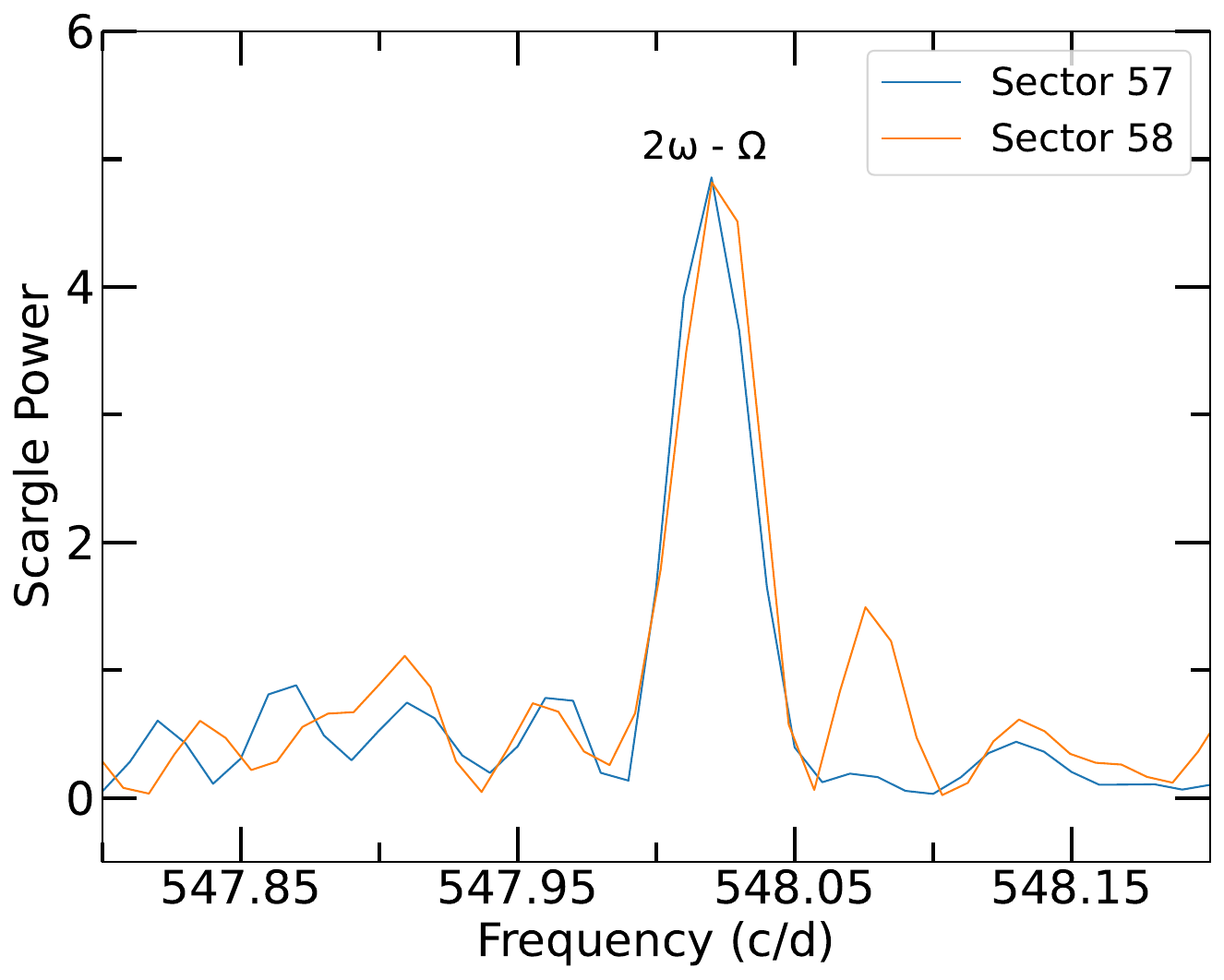}\label{fig:ps_2omega-Omega}}
    \hspace{0.05\textwidth}
    \subfloat[Power spectra of all sectors]{\includegraphics[width=0.47\textwidth]{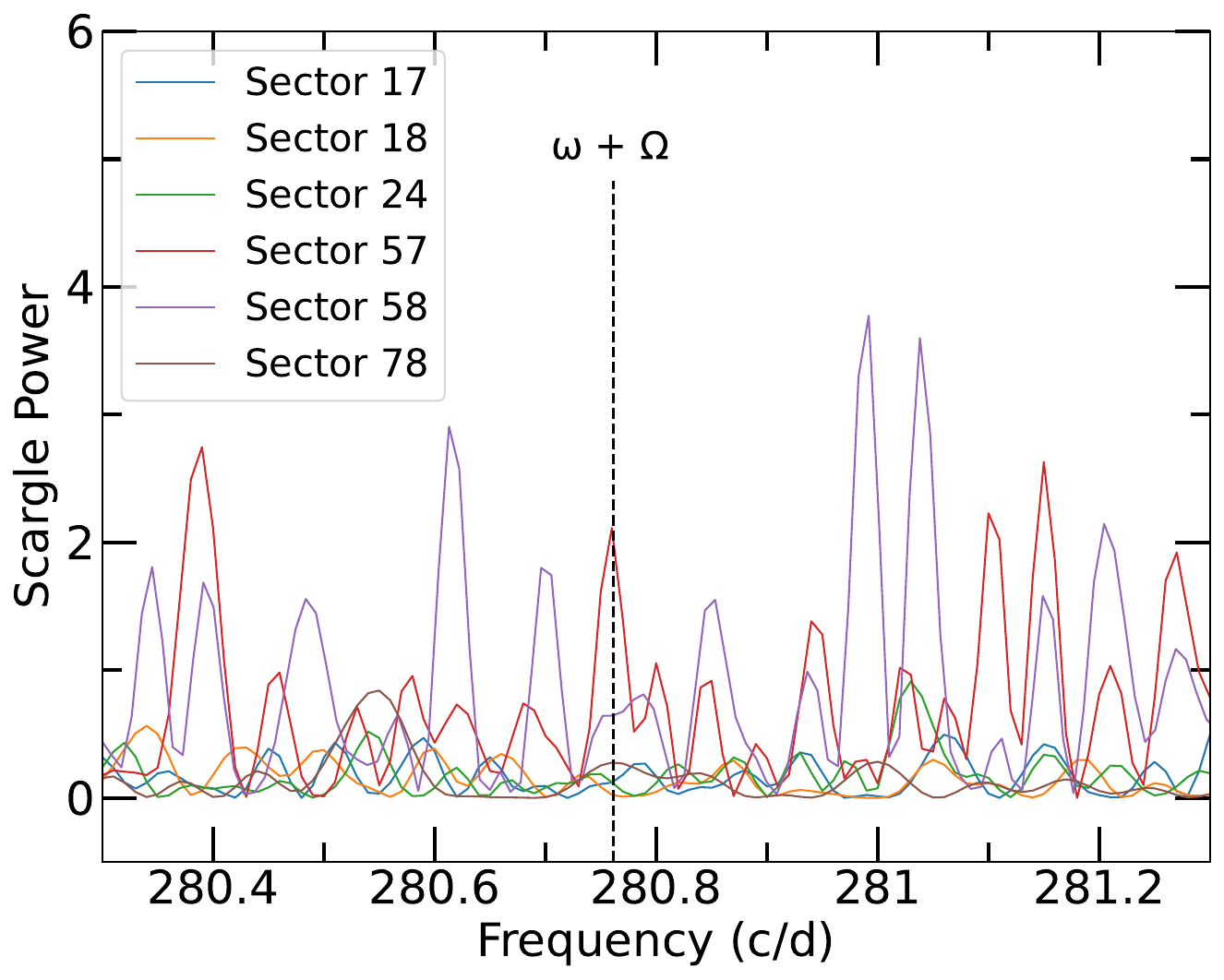}\label{fig:ps_omega+Omega}}
    
    \caption{Power spectra of V709 Cas showing the zoomed-in section near the \twosp - \orb and \spin + \orb.}
    \label{fig:ps_zoomed}
\end{figure*}

In Figure \ref{fig:v709cas_ps}, we observe a prominent peak around the frequency of $\sim$168 cycle d$^{-1}$ for the 120 s cadence dataset; however, this peak is not present in the power spectra for the 20 s cadence data. In the individual sector's power spectra for the 120 s cadence data, this peak was present in all the sectors. Hence, this peak is just a spurious peak that arises due to the cadence. This frequency corresponds to twice the difference between the  Nyquist frequency limit and \spin frequency; thus, it is related to super-Nyquist frequency. We have also closely inspected the high cadence power spectra of 20 s cadence from sectors 57 and 58 and found the signature of the presence of 2\spin-\orb frequency, which is shown in Figure \ref{fig:ps_2omega-Omega}. We have also inspected the presence of \spin+\orb frequency in all power spectra, which is shown in Figure \ref{fig:ps_omega+Omega}. 

\begin{figure}
\centering

    \subfloat[Sector 17]{
       \label{fig:1dps17}
       \includegraphics[width=0.96\textwidth]{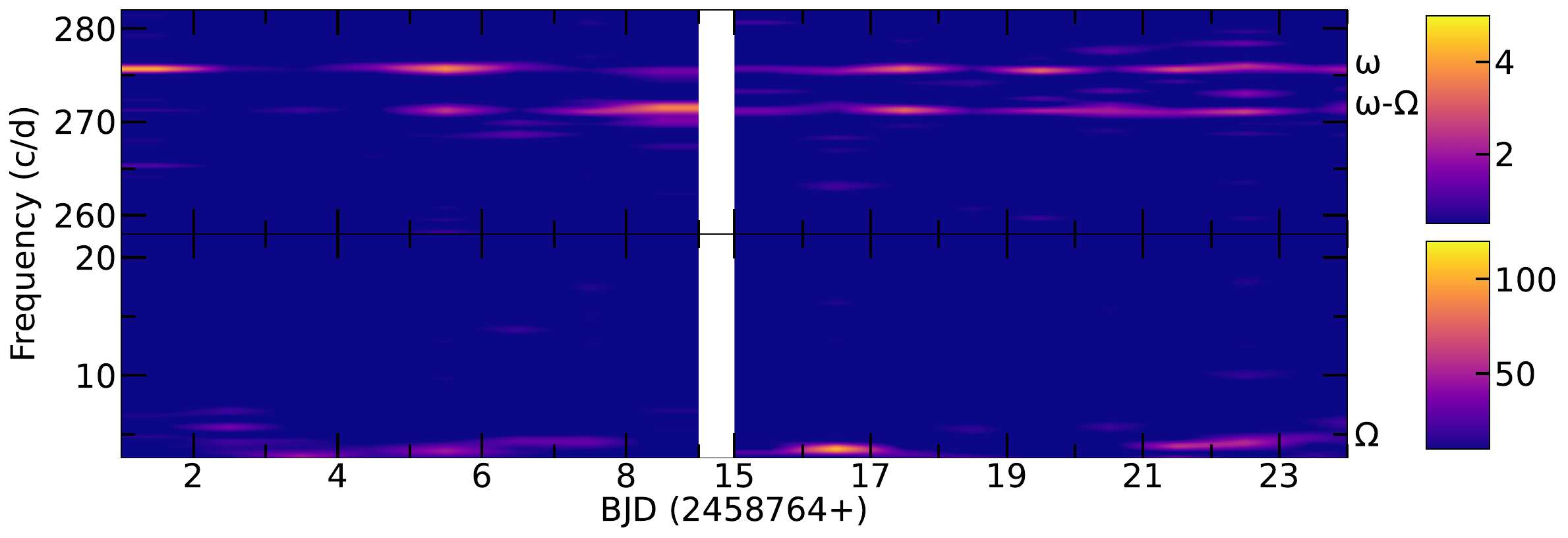}
    }

     \subfloat[sector 18]{
        \label{fig:1dps18}
        \includegraphics[width=0.96\textwidth]{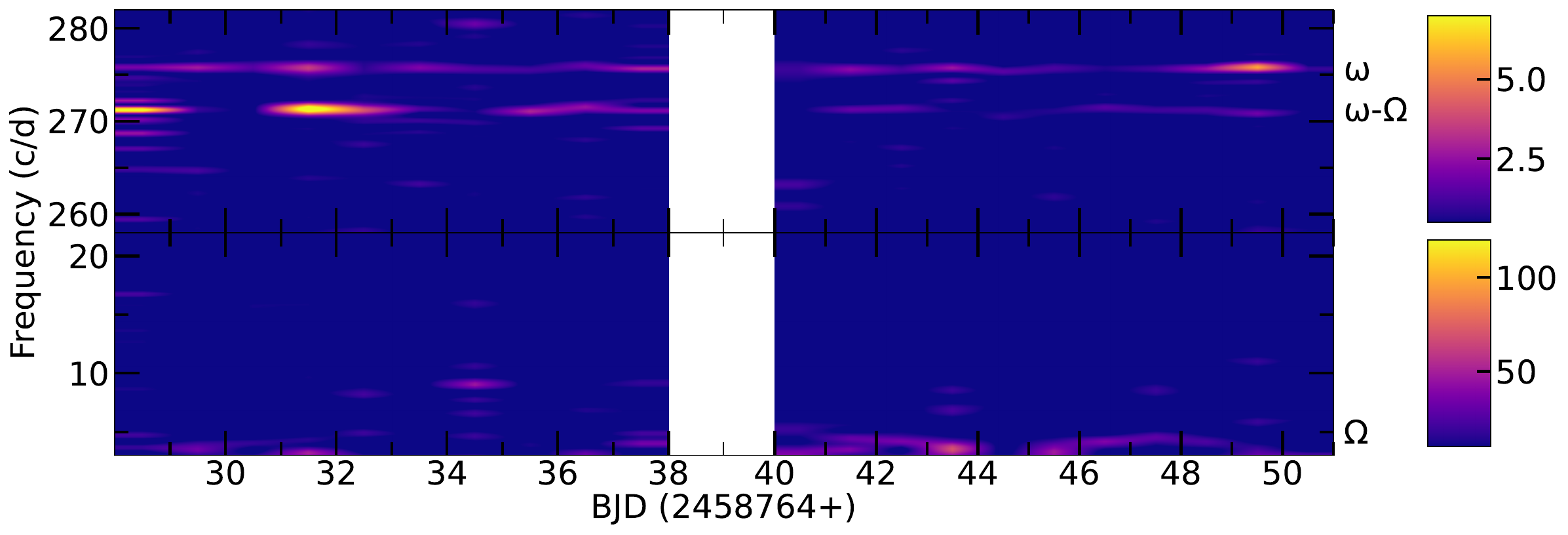}
    }

    \subfloat[sector 24]{
       \label{fig:1dps24}
       \includegraphics[width=0.96\textwidth]{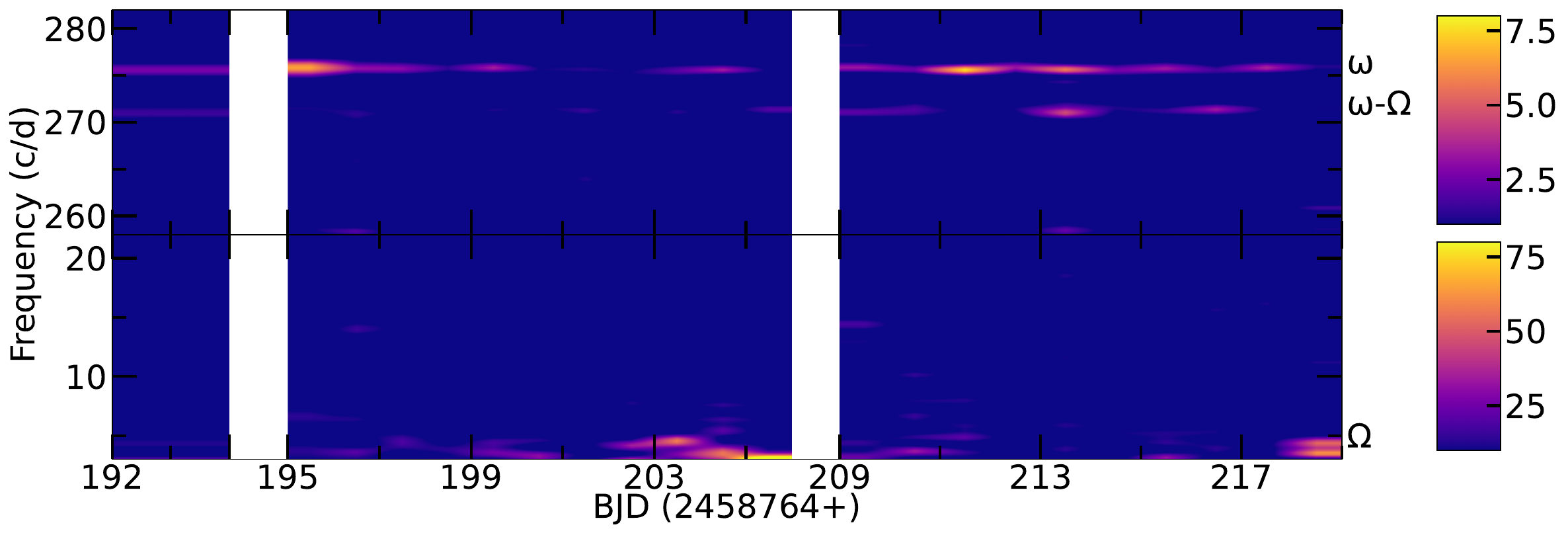}
    }
    \caption{One-day time-resolved power spectra for observations from all sectors using 120 s cadence for the sectors 17, 18, 24 and 78, and 20 s cadence for the sectors 57 and 58. The colour bar on the right side denotes the power.}
    
    \label{fig:1dps_V709_17_18_24}
\end{figure}
 \begin{figure}
 \ContinuedFloat
\centering
    \subfloat[Sector 57]{
        \label{fig:1dps57}

        \includegraphics[width=0.96\textwidth]{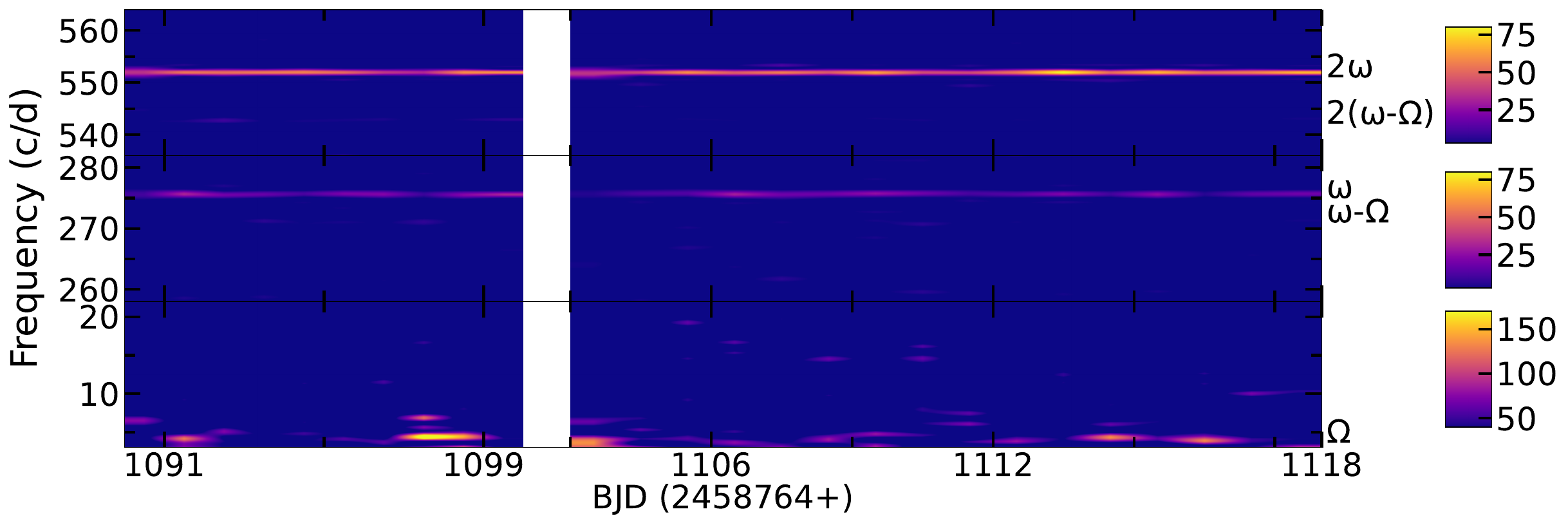}
    }

    \subfloat[Sector 58]{
       \label{fig:1dps58}
       \includegraphics[width=0.96\textwidth]{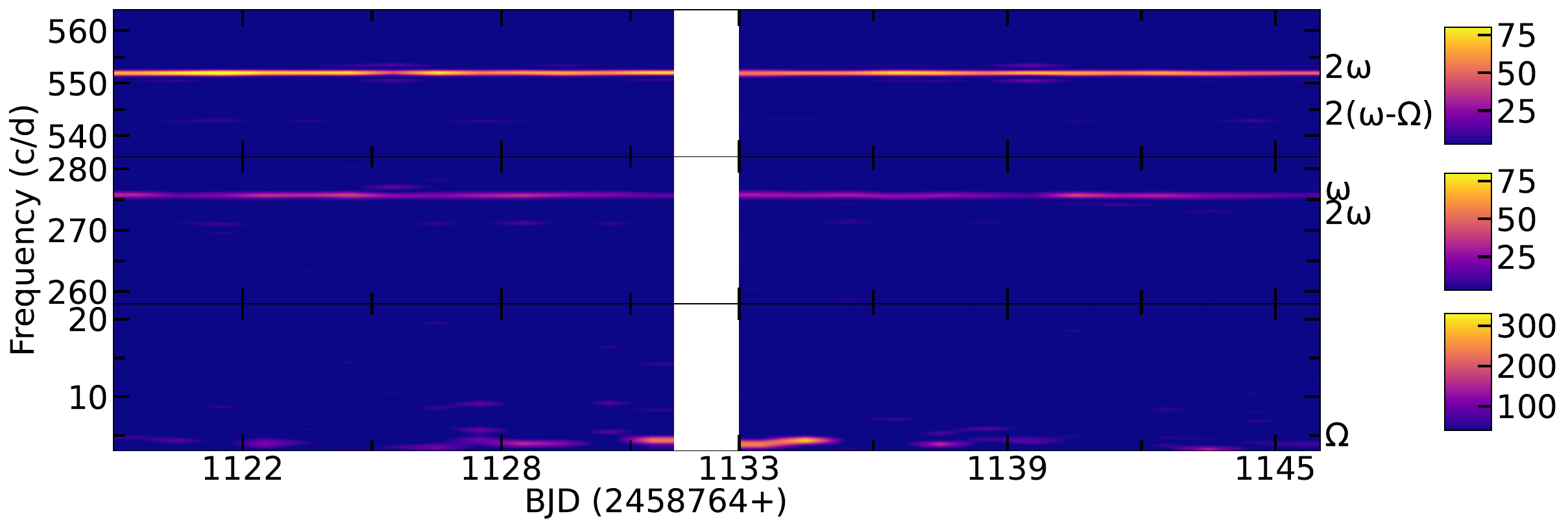}
    }
    
    \subfloat[Sector 78]{
       \label{fig:1dps78}
       \includegraphics[width=0.96\textwidth]{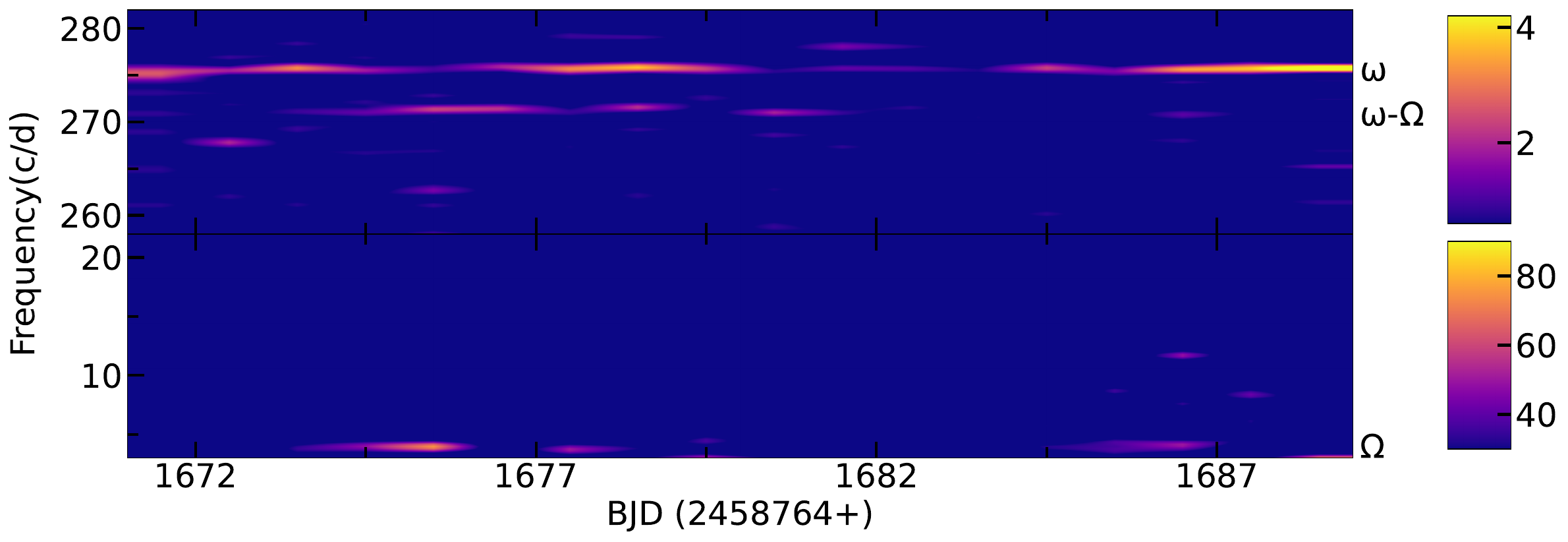}
    }
\caption{Continued ...... }
    \label{fig:1dps_V709}
\end{figure}

\subsection{Time-resolved power spectra} \label{one day}
We have performed the time-resolved power spectral analysis of V709 Cas. The data of all six sectors were divided into one-day time segments, corresponding to more than four orbital cycles. The LS periodogram was performed for each one-day time segment, and time-dependent LS power spectra for all the sectors are shown in Figure \ref{fig:1dps_V709}.  
We employed bilinear interpolation to smooth the time series power spectra for clearer transitions between day-wise power spectra.
The results for sectors 57 and 58 demonstrated similar trends for both the 20 s and 120 s cadence datasets, with the exception of powers on 2\spin and 2\be in the 120 s data.  In the time-resolved power spectra, we obtained \spin, \be, and \orb frequencies for all days with varying power. The \spin frequency was dominant for most days, whereas \be and \orb frequencies were significantly detected only for a few days in sectors 17, 18, 24, and 78. For a few days in sectors 17, 18, 24 and 78, both \spin and \be frequencies had equal dominance.
The short cadence data from sectors 57 and 58 allowed us to obtain \twosp and \twobe, revealing \twosp as the dominant frequency over \spin across all days.
  

\subsubsection{Time evolution of phased light curve and pulse fraction} \label{pf}
\begin{figure*}
    \centering
    \subfloat[120 s Cadence dataset]{\includegraphics[width=\textwidth]{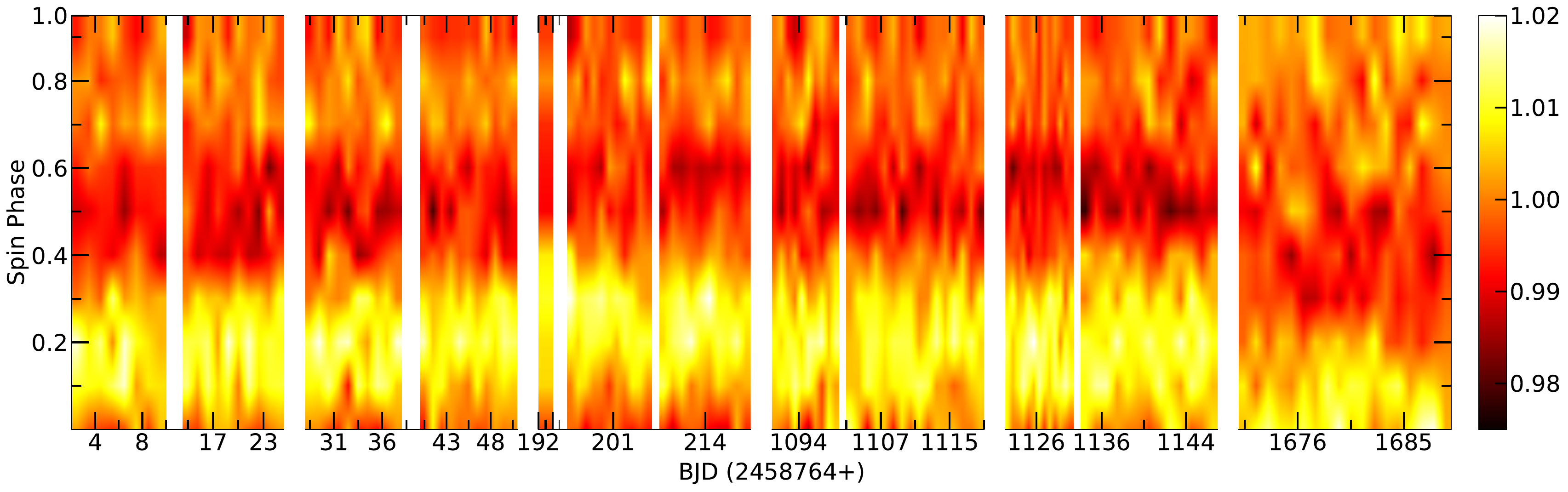}\label{fig:spfold_V709_120}}
    
    \subfloat[20 s Cadence dataset]{\includegraphics[width=\textwidth]{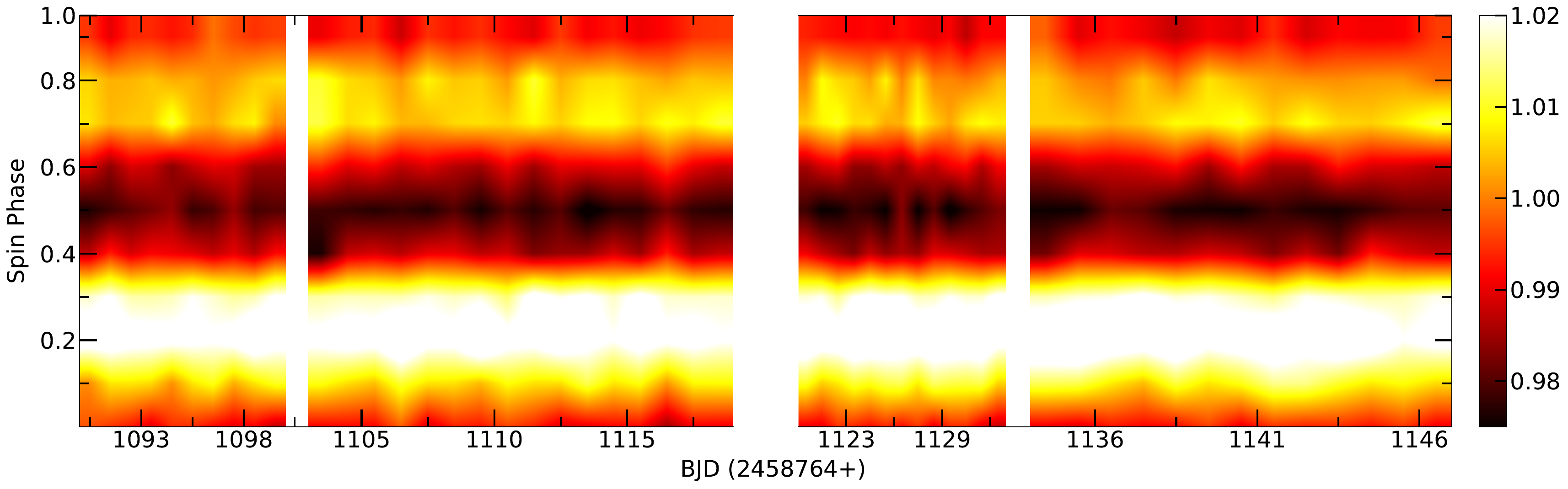}\label{fig:spfold_V709_20}}
    
    \caption{Day-wise spin phase folded light curve of V709 Cas for all sectors' observations. The colour bar on the right side denotes the normalised flux.}
    \label{fig:spfold_V709}
\end{figure*}

We have folded one-day segment light curves for each sector to see the day-wise evolution of the pulse profiles using our derived \psp. The time of the first data point from the \textit{TESS} observations was taken as an epoch of zero phase. For sectors 57 and 58, we have folded a one-day time segment light curve for both cadences, 120 s and 20 s. The spin pulse profile structure for all the sectors for the 120 s cadence dataset is shown in Figure \ref{fig:spfold_V709_120} and that for the 20 s cadence dataset is shown in Figure \ref{fig:spfold_V709_20}. The folded light curves obtained with a 20 s cadence provide a more defined spin pulse profile than those obtained with a 120 s cadence.
In all the sectors, two peaks in the folded light curve were seen, with one peak having higher flux than the other. The first peak, with a higher amplitude, was located around the 0.25 phase, while the second peak was situated near the 0.75 phase, resulting in a phase separation of approximately 0.5. In sector 78, the two-peak structure was less discernible.

\begin{figure*}
    \centering
    \includegraphics[width=\textwidth]{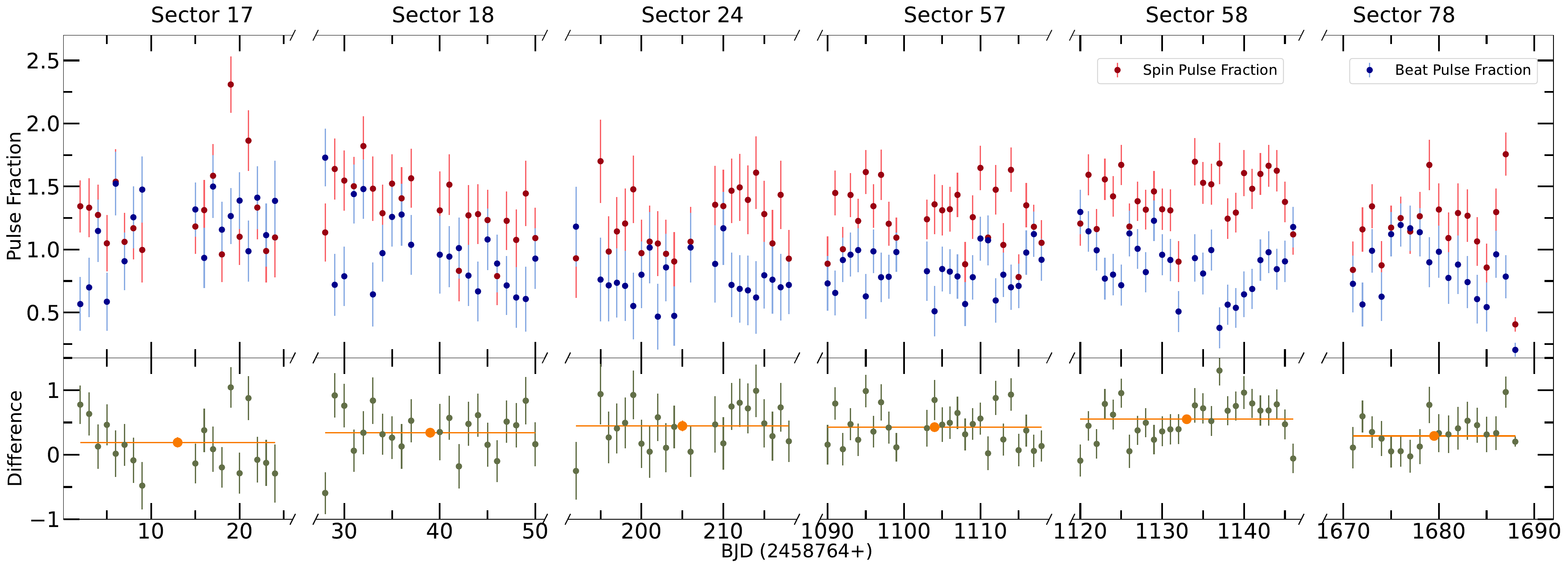}
    \caption{The top panel shows the spin and beat pulse fraction of all sectors and the bottom panel shows the difference between the spin and beat pulse fraction on each day. The orange solid circles show the mean difference between the spin and beat pulse fractions of the individual sector, and the error bar on the x-axis shows that sector’s start and end times.}
    \label{fig:pf_V709}
\end{figure*}

We have also calculated the pulse fraction for the spin and beat-phased light curve of each one-day segment as 
\begin{equation}
    P.F = \frac{A_{max} - A_{min}}{A_{max} + A_{min}} \times 100 \% 
\end{equation}
Here, $A_{max}$ and $A_{min}$ are the maximum and minimum flux values of each light curve, respectively. These derived spin and beat pulse fractions were plotted against the corresponding days and shown in the top panel of Figure \ref{fig:pf_V709}. The bottom panel represents the difference between the spin and beat pulse fractions. We see that for most of the days, this difference is positive. However, during sectors 17, 18, and 78, we do not see a significant difference between the spin and beat pulse fractions, whereas, during sectors 24, 57, and 58, the difference in both became relatively larger than the earlier sectors, with the spin pulse fraction being slightly higher than the beat pulse fractions. The spin pulse fraction was found to be a maximum of $2.3 \pm 0.2 $ \%, whereas the beat pulse fraction was found to be $1.7 \pm 0.2$ \%. The minimum values of spin and beat pulse fractions were found to be $0.40 \pm 0.05$ \% and $0.20 \pm 0.05$ \%, respectively. 


\section{Discussion}
\label{sec:dis}
Detailed timing analysis of IP V709 Cas has been performed using high-cadence optical photometric data acquired by the \textit{TESS} over an extended period. The \porb, \psp, and \pbe are determined to be $5.3329 \pm 0.0002$ h, $312.7478 \pm 0.0002$ s and $317.9267 \pm 0.0002$ s, respectively, and are derived with a better precision. These periods are also consistent with those reported in earlier studies  \citep{Motch1996, Norton1999, Kozhevnikov2001, Thorstensen2010}. The accretion mechanisms can be explained based on the presence of frequencies in the power spectra \citep{Ferrario1999,Wynn1992}. The disc-fed accretion leads \spin frequency modulations \citep{1996A&A...307..824K}, while stream-fed accretion results in modulations at the lower orbital sideband, \be \citep{Ferrario1999}. The disk-overflow accretion is predicted to show modulations at both \spin and \be, distinguished by their relative amplitudes \citep{Hellier1991, Hellier1993}. In the majority of the sectors, the \orb frequency is found to be dominant; however, the \twosp was found to be dominant for the short cadence data of sectors 57 and 58. Dominance of \orb frequency could be due to amplification of \orb power by the modulation of \spin frequency with \be, i.e. \spin - (\be) = \orb. If this is the case, then this should also produce an equal power at the upper sideband too, i.e. \spin + (\be) = 2\spin - \orb \citep{Warner1986}. We could not detect the  2\spin-\orb frequency as this frequency falls outside the Nyquist frequency of the 120 s cadence data but, in the 20 s cadence data of sectors 57 and 58, we could see this frequency as shown in Figure \ref{fig:ps_2omega-Omega}. Similarly \be frequency could also arise because of the modulation of the \spin and \orb frequencies. However, no peak was observed at \spin + \orb in any of the sectors' power spectra. This suggests that the modulation in the \be frequency is intrinsic, demonstrating that some accretion gets through the streams in V709 Cas. However, the dominance of \spin over the \be suggests that the system is a disc-overflow accreator with disc dominance accretion \citep{Hellier1998}. There is a significantly higher power at \twosp and  \twobe frequencies for sectors 57 and 58. In a pure stream-fed system, if significant power is observed at the \twobe then the magnetic field distribution of the WD possesses an up-down field symmetry with respect to the orbital plane \citep{Ferrario1999}. Since a part of the accretion appears to be stream-fed in V709 Cas; therefore,  \twobe frequency can be attributed to the up-down field symmetry. A higher power at \twosp suggests that there is a substantial contribution from both magnetic poles of the WD \citep{Szkody2017}, which reinforces the previously suggested two-pole accretion geometry of V709 Cas \citep{Norton1999}.

Time-resolved power spectral analysis indicates that the system predominantly exhibits characteristics consistent with disc-overflow accretion. In most observational days, the power associated with the \spin frequency is more than that of the \be frequency. However, in a subset of observations, the \be frequency is either dominant or displays comparable power to the \spin frequency, with minimal distinction in their respective power levels. The changing accretion geometry is more prominent in sectors 17, 18, 24, and 78, whereas in sectors 57 and 58, the accretion geometry appears to be always disc-fed dominant and is not varying. For most of the days in sectors 17, 18, 24, and 78, the accretion geometry is disc-fed dominant, and for only a few days, it is either stream-fed dominant or both disc and stream-fed have equal dominance. For example, for the days 8, 9, 15, 20, 22, 28, 31, 32, 35, 36, 42, 46, 201, 206, 216, 1675, and 1680, the power of \be frequency dominates \spin frequency power, whereas for days 15, 17, 24, 47, 218, 1676 powers of \be and \spin are equal. This observed change could be due to variations in the mass accretion rate \citep{deMartino1995, Buckley1996, deMartino1999}. An increased mass accretion rate could potentially cause more material to overflow from the disc and directly hit the magnetosphere, leading to a stream-fed dominance. When the mass accretion rate decreases, then the amount of overflow material decreases, and we observe a disc-fed dominance. In sectors 57 and 58, the mass accretion rate is non-varying, which can be observed from the relatively constant mean flux of individual days (see Figure \ref{fig:mean_lc}); hence, no change in the dominant accretion flow is noticed throughout these two sectors' observations. In the remaining sectors, the mass accretion rate appears to be variable as indicated by the daily mean flux variations in Figure \ref{fig:mean_lc}. Thus, switching the dominance between disc-fed and stream-fed accretion is noticed. This type of switching of the accretion mode is also observed in other IPs such as TX Col \citep{Rawat2021,Littlefield2021}, FO Aqr \citep{Littlefield2020} and V902 Mon \citep{Rawat2022}. The IP, V2400 Oph was also found to be changing powers of \spin and \be frequencies,  and it is suggested that this change in the dominance could be due to the variation in the density and distribution of the diamagnetic blobs, which influence whether accretion occurs in a disc-fed or stream-fed manner \citep{Joshi2019,Langford2022}. Furthermore, in the case of  V709 Cas, no discernible secular trends were observed in the variations of powers of these frequencies, precluding definitive conclusions regarding a potential evolution of the accretion geometry. The accretion in V709 Cas appears to exhibit time-dependent variations, transitioning between periods of stream-dominated and disc-dominated of the beat pulse fraction. However, during sectors 17, 18, 24, and 78, we do not see a significant difference in accretion in a short time, resulting in a dynamically complex accretion environment.

The presence of double-humped \psp modulations in the spin-phase folded light curves further supports our claim of it being a two-pole accretor. The interplay between the dominance of spin and beat modulations is also consistent with the pulse-fraction values obtained from the time-resolved phased light curves of V709 Cas. For most days, the difference stays within the 1$\sigma$ level. However, in sectors 24, 57, and 58, the spin pulse fraction is usually higher than the beat pulse fraction, indicated by a positive difference. This consistently larger difference suggests that disc-fed accretion is more dominant in these specific sectors.

The mean flux of individual sectors appears to be changing among the sectors as shown in Figure \ref{fig:mean_lc}. This result supports the earlier findings by \cite{Motch1996} and \cite{deMartino2001} of change in luminosity in a time scale of a month. The increase in luminosity might be due to the increased mass accretion rate. The magnetospheric radius depends upon the mass accretion rate ($\propto \Dot{M}^{-2/7}$). Thus, an increase in the mass accretion rate will decrease the magnetospheric radius, resulting in the accretion onto the magnetic poles over a wide area. The increased amount of material falling onto the magnetic poles absorbs more radiation in the vertical direction \citep{deMartino2001}. This increases the amplitude in the spin phase folded light curves, resulting in higher spin pulse fraction values. Please note that comparing the TESS flux among the different sectors' observations may require calibrations with ground-based observations \citep[see e.g.][]{Scaringi2022}. We could not calibrate the \textit{TESS} data with the available ground-based observations (e.g. ASAS-SN; All Sky Automated Survey for Supernovae g-band)  due to insufficient simultaneous observations.


\section{Summary}
\label{sec:conc}
The first long-term, short cadence and continuous TESS observations of V709 Cas have yielded precise orbital (5.3329 $\pm$ 0.0002 h), spin (312.7478 $\pm$ 0.0002 s), and beat (317.9267 $\pm$ 0.0002 s) periods. Analysis of power spectra from various sectors indicates the presence of both spin and beat frequencies, with the spin frequency exhibiting greater power in most cases. The short cadence observations enabled us to do day-wise time-dependent timing analysis, which also revealed similar results. These results are also supported by the time evolution of spin and beat-phased light curves, where the spin pulse fraction was found to be more than the beat pulse fraction for the majority of the days.  The bulk of power spectral analysis and phased light curve analysis consistently indicate that V709 Cas is a disc-overflow system with disc-fed accretion as the primary mechanism; however, stream dominance is also seen occasionally.

\section*{Acknowledgement}
We thank the referee for reading our manuscript and providing useful comments and suggestions. This paper includes data collected by the TESS mission. Funding for the TESS mission is provided by the NASA's Science Mission Directorate.


\bibliographystyle{elsarticle-harv} 
\bibliography{references}

\end{document}